\documentclass[10pt,letterpaper,conference]{IEEEtran}
\IEEEoverridecommandlockouts
\usepackage{cite}
\usepackage{amsmath,amssymb,amsfonts}
\usepackage{algorithmic}
\usepackage{graphicx}
\usepackage{textcomp}
\usepackage{xcolor}

\usepackage{color,soul}

\def\BibTeX{{\rm B\kern-.05em{\sc i\kern-.025em b}\kern-.08em
    T\kern-.1667em\lower.7ex\hbox{E}\kern-.125emX}}

\begin{document}

\title{Hardware implementation of digital \\ memcomputing on small-size FPGAs
\thanks{This work was supported by the NSF grant No. ECCS-2229880.}
}


\author{\IEEEauthorblockN{Dyk Chung Nguyen\IEEEauthorrefmark{1},
Yuan-Hang Zhang\IEEEauthorrefmark{2}, 
Massimiliano Di Ventra\IEEEauthorrefmark{2}\IEEEauthorrefmark{3} and
Yuriy V. Pershin\IEEEauthorrefmark{1}\IEEEauthorrefmark{3}}
\IEEEauthorblockA{\IEEEauthorrefmark{1}Department of Physics and Astronomy\\
University of South Carolina,
Columbia, South Carolina 29208, USA}
\IEEEauthorblockA{\IEEEauthorrefmark{2}Department of Physics, University of California, San
Diego, La Jolla, CA, 92093-0319, USA}
\IEEEauthorblockA{\IEEEauthorrefmark{3}
Email: diventra@physics.ucsd.edu; pershin@physics.sc.edu}
}
\maketitle

\begin{abstract}
Memcomputing is a novel computing paradigm beyond the von–Neumann one. Its digital version is designed for the efficient solution of  combinatorial optimization problems, which emerge in various fields of science and technology. Previously, the 
performance of digital memcomputing machines (DMMs) was demonstrated using software simulations of their ordinary differential equations. Here, we present the first {\it hardware} realization of a DMM algorithm on a low-cost FPGA board. In this demonstration, we have 
implemented a Boolean satisfiability problem solver. To optimize the use of hardware resources, the algorithm was partially parallelized. The scalability of the present implementation  is explored and our FPGA-based results are compared to those obtained using a python code running on a traditional (von–Neumann) computer, showing one to two orders of magnitude speed-up in time to solution. This initial small-scale implementation is projected to state-of-the-art FPGA boards anticipating further advantages 
of the hardware realization of DMMs over their software emulation.
\end{abstract}


\section{Introduction}

During the past decade or so, the evolution of conventional computing devices has slowed down prompting the research community to shift its focus to alternative computing paradigms, such as neuromorphic, quantum, and stochastic computing, to name just a few. In many of such approaches, the computation is implemented beyond the von-Neumann paradigm, thus bypassing some of the most 
pressing issues (the von Neumann bottleneck, energy consumption, and further scaling) of our computing technology. It is then hoped that unconventional computing may boost or even revolutionize computing~\cite{Finounc}.

One such alternative approach is {\it memcomputing}~\cite{diventra13a,UMM,memcomputingbook}, which relies on the ability of some physical systems 
to exploit memory (time non-locality)~\cite{09_memelements} to process information directly on the same physical platform where the computation result is ultimately stored. It is important to note here that time non-locality (e.g., the property that when the state of a physical system is perturbed, the perturbation affects the system’s state at a later
time~\cite{memcomputingbook,Membook}) is an essential feature to the computation. This is because time non-locality promotes spatial non-locality in the system~\cite{memcomputingbook}. This spatial non-locality, in the form of {\it dynamical long-range order}, can then be exploited to solve computational problems efficiently. In this approach, then the massively-parallel dynamics of these physical systems at a dynamical long-range ordered state is essentially the process of computation. As the dynamics of such systems can be described by coupled ordinary differential equations (ODEs), finding the equilibrium point(s) of such a set of coupled ODEs is then equivalent to the physical computation~\cite{memcomputingbook,SIEGELMANN1998214}.  
In particular, the digital memcomputing machines (DMMs) solving combinatorial optimization problems~\cite{Traversa17a} can be designed so that their phase space has a single attractor corresponding to the problem solution (or multiple attractors if several solutions are possible). Additionally, the dynamics of these machines is deterministic, without periodic orbits or chaos~\cite{no-chaosa,no-chaosb}, and topologically robust against perturbations and noise~\cite{topo,DMtopo}.

So far, all applications of DMMs have been based on software simulations of their ODEs, see, e.g., Refs.~\cite{memcomputingbook,stress-test,Sean3SAT}. However, the hardware realization has well-recognized benefits including better scalability and faster execution time. Indeed, in contrast to prior work wherein DMMs have been simulated in a sequential fashion, in hardware, the DMM algorithms may be implemented in parallel. Therefore,  we expect the hardware implementation of DMMs to provide additional benefits.
In this work, we implement a 3-SAT solver using an entry-level FPGA board. Our goal is to understand the advantages and limitations of this implementation and to develop optimization strategies for the future deployment of memcomputing technologies on larger boards.

We emphasize that memcomputing is a {\it classical} (non-quantum) approach to computation in the sense that it does not rely on quantum features such as entanglement. Although ``quantum supremacy'' over traditional computers has been claimed recently using a superconducting quantum information processor~\cite{arute2019quantum}, it still remains unclear how this demonstration can be transformed into a useful computation. At the same time, there are solid results on the advantages in scalability offered by memcomputing in the solution of combinatorial optimization problems (see, e.g., Ref.~\cite{memcomputingbook,memcpu}). 

\begin{figure*}[t]
\centering
    \includegraphics[width=0.8\textwidth]{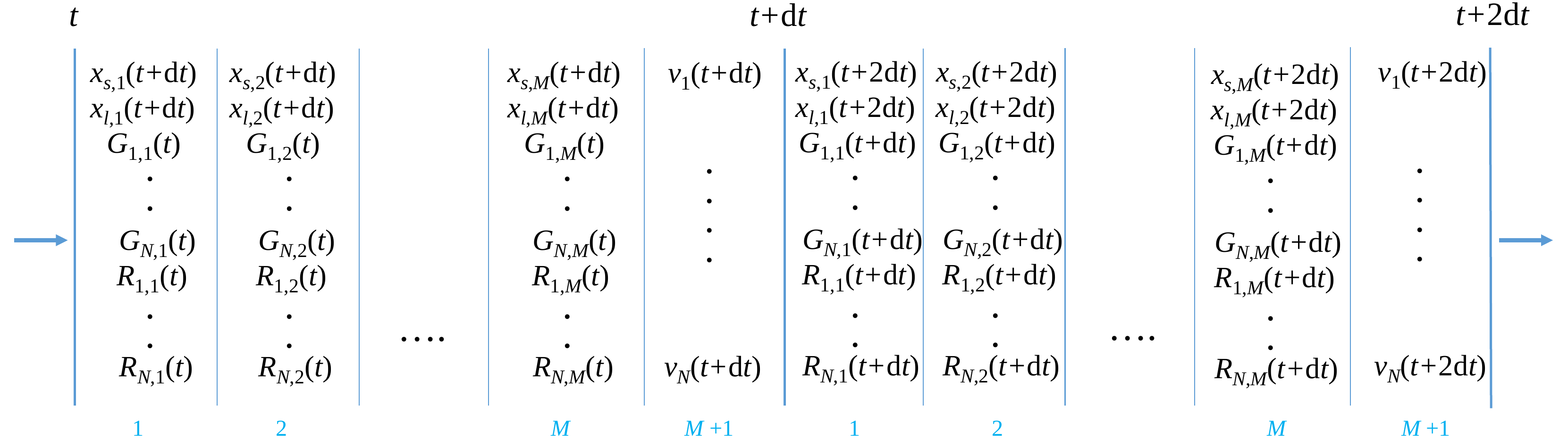}
    \caption{Schematics of the parallerization of the memcomputing model. See the text for details.}
    \label{fig:1}
\end{figure*}

This paper is organized as follows. In Sec.~\ref{sec:2}, we briefly present the 3-SAT problem and memcomputing ODEs used in this work. Some related work and approaches are summarily given in Sec.~\ref{sec:22}. Next, we provide details of the FPGA implementation, which we compare with the performance of a code written in python to solve those ODEs (Sec.~\ref{sec:3}). Our results are presented in Sec.~\ref{sec:4} and discussed in Sec.~\ref{sec:5}. Finally, we offer our 
conclusions in Sec.~\ref{concl}.

\section{memcomputing equations for 3-SAT} \label{sec:2}
3-SAT is a particular case of satisfiability (SAT) problems, wherein each clause is the disjunction of 3 literals. A literal is a Boolean variable or its negation. The problem consists in finding the assignment of variables such that all clauses evaluate to true.

To tackle the 3-SAT, we have selected the DMM equations introduced in Ref.~\cite{Sean3SAT}. These equations are:

\begin{eqnarray}
\dot{v}_n&=&\sum\limits_mx_{l,m}x_{s,m}G_{n,m}(v_n,v_j,v_k)+\left( 1+\zeta x_{l,m}\right)\cdot \nonumber \\
& & \left(1-x_{s,m}\right) R_{n,m}(v_n,v_m,v_k), \label{eq:1}\\
\dot{x}_{s,m}&=&\beta \left( x_{s,m} +\epsilon \right)\left( C_m(v_i,v_j,v_k)-\gamma\right),\\
\dot{x}_{l,m}&=&\alpha \left( C_m(v_i,v_j,v_k)-\delta\right),\\
G_{n,m}&=&\frac{1}{2}q_{n,m}\text{min}\left[\left( 1-q_{j,m}v_j\right), \left( 1-q_{k,m}v_k\right)\right],\\
R_{n,m}&=&\begin{cases}
    \frac{1}{2}\left(q_{n,m}-v_n \right), \\
    \hspace{1cm} \text{if } C_m(v_n,v_j,v_k)=\frac{1}{2}\left(1-q_{n,m}v_n \right),\;\;\\
    0, \hspace{7mm} \text{otherwise}.
  \end{cases}
\end{eqnarray}
Here, $v_n$ are continuous variables ($n=1,\ldots,N$),  $x_{s,m}$ and $x_{l,m}$ are the memory variables ($m=1,\ldots,M$), where $N$ is the number of variables, $M$ is the number of clauses, $q_{j,m}=1$ if the $j$-th variable enters $m$-th clause, $q_{j,m}=-1$ if the negation of the $j$-th variable enters $m$-th clause. Moreover, $\alpha$, $\beta$, $\gamma$, $\delta$, $\epsilon$ and $\zeta$ are constants~\cite{Sean3SAT}. We emphasize that in the above equations, each variable of the 3-SAT is represented by a continuous quantity $v_n$, and two memory variables, the short (s) and long (l), are associated with each clause. Additionally, $v_n$-s are constrained to the interval $[-1,1]$, $x_{s,m}$-s are constrained to the interval $[\epsilon, 1-\epsilon]$, and $x_{l,m}$ to the interval $[1,10^4M]$. The sign of $v_n$ defines its Boolean value: 0 if the sign is negative, and 1 otherwise.

Moreover, the clause function $C_m(v_i,v_j,v_k)$ is defined as
\begin{align}
& C_m(v_i,v_j,v_k)= \nonumber \\     
 &  \hspace{3mm}  \frac{1}{2}\text{min}\left[\left(1-q_{i,m}v_i \right),\left(1-q_{j,m}v_j \right),\left(1-q_{k,m}v_k \right)\right].\label{eq:6}
\end{align}
This function characterizes the state of the variable that most closely 
satisfies the clause $m$. For more information on the model, and how it has been obtained, see Refs.~\cite{Sean3SAT,memcomputingbook}.

\section{Related work} \label{sec:22}

The integration of ODEs on FPGAs and hardware acceleration of SAT were explored in the past. For instance, Stamoulias {\it et al.}~\cite{stamoulias2017high} reported a high-performance FPGA accelerator for the hardware integration of ODEs that achieves up to 14x speedup compared to single-core CPU solution~\cite{stamoulias2017high}. Recently, Hollabough and Chakraborty~\cite{9937835} programmed an FPGA co-processor to solve Lotka-Volterra equations. They reported 4.8x speedup. 

Moln\'{a}r {\it et al.}~\cite{molnar2020accelerating} used GPUs to accelerate a continuous-time analog SAT solver. They reported up to two orders of magnitude improvement compared to the CPU implementation. 
Sohanghpurwala {\it et al.}~\cite{sohanghpurwala2017hardware} published a survey of hardware-accelerated SAT solvers. For a comprehensive review of various hardware accelerators in general see Ref.~\cite{peccerillo2022survey}.
In view of these previous results, one can then expect a similar speed-up of memcomputing on FPGAs.

\section{Methods} \label{sec:3}

\subsection{Generation of SAT instances}


To generate random ``hard'' 3-SAT instances, we use the algorithm proposed in \cite{barthel2002hiding}. First, a planted solution is set to $y_i=1$ for all $i\in\{1, \ldots, N\}$, where $y_i$ is the $i$-th Boolean variable. Then, we generate $M$ clauses randomly and independently. Each clause has three literals randomly selected from the $N$ variables and can have either 0, 1, or 2 negated literals, with probabilities $p_0$, $3p_1$, and $3p_2$, respectively. The factor of 3 accounts for the fact that there are 3 possible ways to negate either 1 or 2 literals. Clauses with 3 negated literals are excluded since $\bar{y}_i \vee \bar{y}_j \vee \bar{y}_k$ is not satisfied under the planted solution.

To ensure that the generated 3-SAT instances are hard, the probabilities should satisfy the following constraints:
\begin{equation}
0.077 < p_0 < 0.25,\ p_1 = \frac{1 - 4 p_0}{6},\ p_2 = \frac{1 + 2 p_0}{6}.
\end{equation}
The proof of these constraints can be found in \cite{barthel2002hiding}. In this work, we choose $p_0=0.08$. 

Finally, we generate a random planted solution by randomly negating each variable $y_i$ with probability 0.5 and negating all occurrences of $y_i$ in all clauses.

\subsection{FPGA implementation} 

In this work, we used a commercially available Nexys A7-100T Artix-7 FPGA board  from Digilent (FPGA part XC7A100T-1CSG324C). This FPGA contains 63,400 lookup tables (LUTs), 15,850 logic slices, and 126,800 CLB flip-flops. The clock speed is 100~MHz. To program the FPGA we used Verilog, which is a hardware design language (HDL). Its syntax is based on C programming language, making it fast and easy to program.

There are multiple ways to implement Eqs.~(\ref{eq:1})-(\ref{eq:6}) in FPGAs. Although the entire parallelization of integration steps in Eqs.~(\ref{eq:1})-(\ref{eq:6}) is possible\footnote{In this case, all $M+1$ steps in Fig.~\ref{fig:1} are merged into a single step.}, such an approach is highly resource-demanding and, therefore, not useful in practice. In fact, the largest problem that we were able to implement fully in parallel with these small-size FPGAs contained 3 variables and 6 clauses. 

Therefore, we have adopted a hybrid approach, see Fig.~\ref{fig:1} for the general scheme. Within this approach,
the time period between $t$ and $t+\textnormal{d}t$ is divided into $M+1$ intervals. The first $M$ intervals are used to evaluate variables and functions related to the $m$-th clause and compute the sums in the right-hand side of Eq.~ (\ref{eq:1}) (not shown in Fig.~\ref{fig:1}).
The last step in the sequence is used to update the voltages based on the sums evaluated in the previous steps. Essentially, this procedure corresponds to the forward Euler integration at a constant time step.

We note that this implementation relies on trading a clock cycle on computing $C_m$, $G_{n,m}$, $R_{n,m}$ for each clause by using block RAM of Nexys A7-100T. The last $(M+1)$-th step is also needed to implement a delay between reading and writing variables into the block RAM of FPGA.

\subsection{Python code}
To compare with the FPGA implementation, we also implemented the same equations using PyTorch \cite{paszke2019pytorch}. The code is publicly available online at \cite{github_link}. 

\begin{figure}[t]
\centering
    (a) \includegraphics[width=0.45\textwidth]{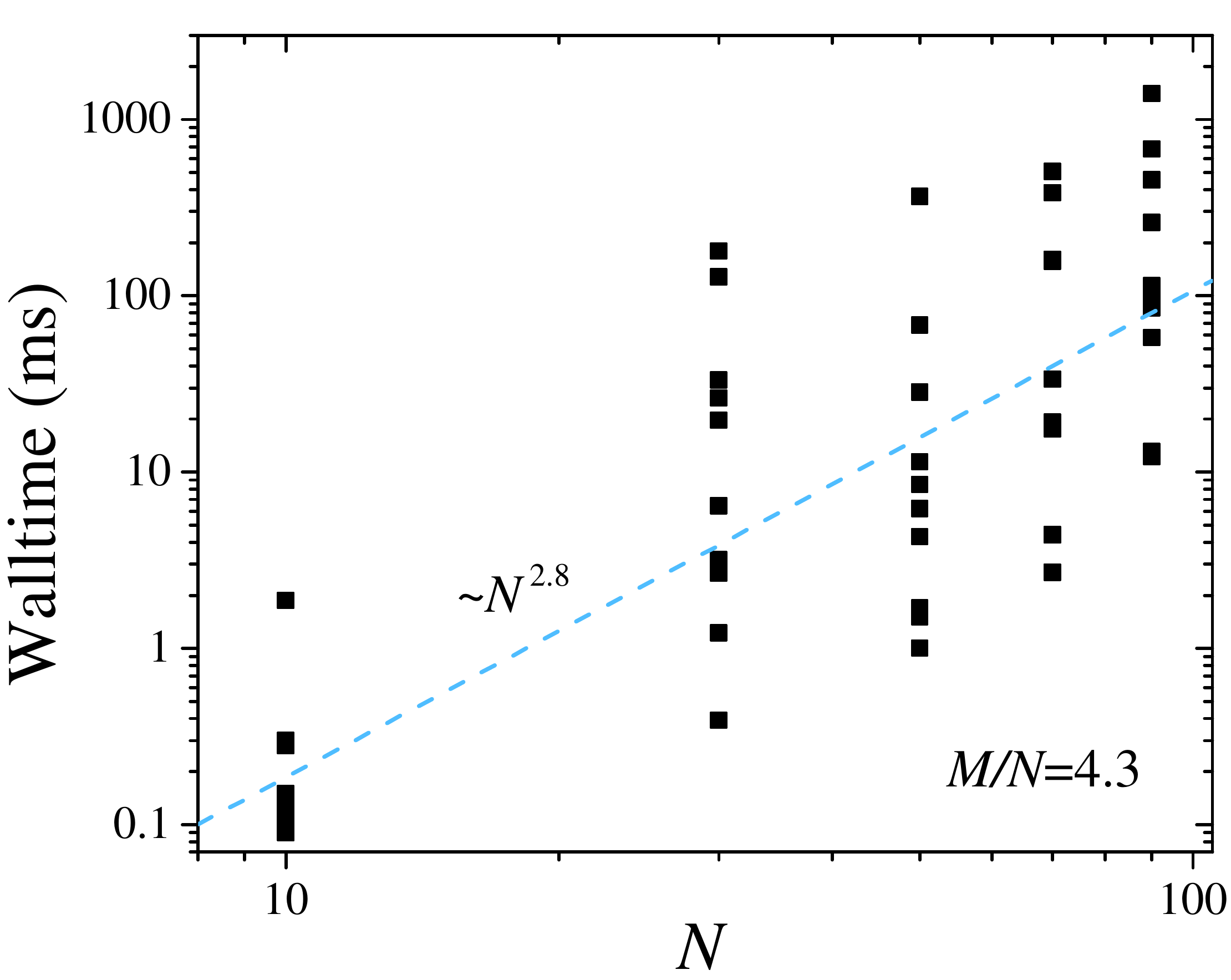} \\
    (b) \includegraphics[width=0.45\textwidth]{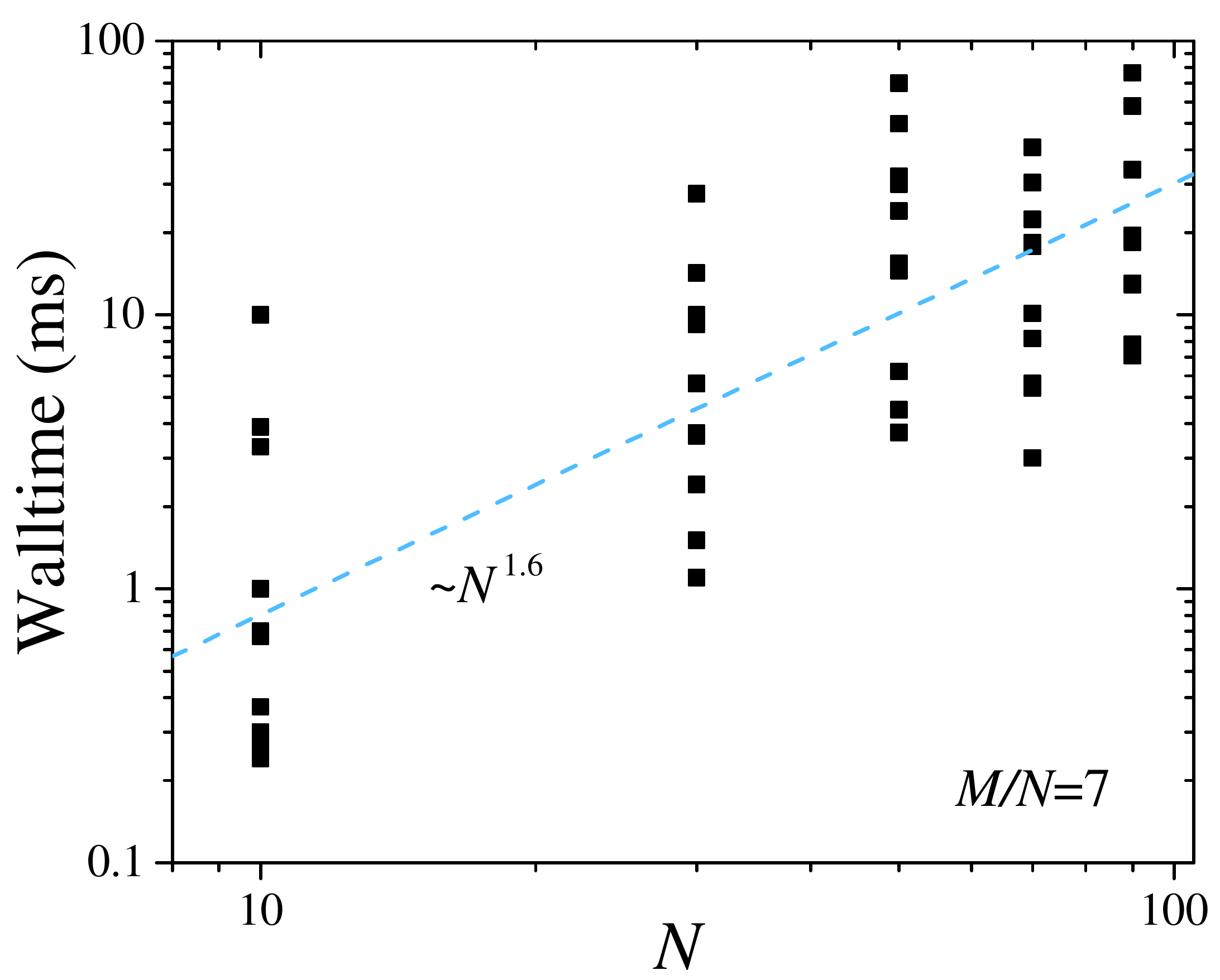} 
    \caption{Time to reach the solution, $T_\textnormal{FPGA}$, as the function of the number of variables for (a) $M/N=4.3$ and (b) $M/N=7$ using FPGA board. Median times were used to obtain the fits (see the text).}
    \label{fig:2}
\end{figure}

\section{Results} \label{sec:4}

The memcomputing approach was implemented in hardware for $M/N=4.3$ and $7$, $p_0=0.08$, and $N=10$, $30$, $50$, $70$, and $90$. Instances are typically more difficult on approaching $M/N\sim 4.3$~\cite{Mezard}.  
For each value of $N$, we generated 10 instances of 3-SAT that were used (without pre-selection) to compile the FPGA code. The initial conditions utilized in FPGA and python  calculations were the same. Specifically, we used randomly generated initial values of variables (in the interval from -1 to 1), $x_{l,m}(0)=1$, $x_{s,m}(0)=C_m(0)$ that are the initial values of corresponding clause functions $C_m$.

The results presented herein were obtained using the following set of parameter values: $\alpha=5$, $\beta=20$, $\gamma=1/4$, $\delta=1/20$, $\epsilon=10^{-3}$, $\zeta=0.1$ for $M/N = 7$, and $\zeta=0.001$ for $M/N=4.3$~\cite{Sean3SAT}.
In all cases considered in this work, the problem solutions were found. 
Fig.~\ref{fig:2} shows the resulting calculation times for $M/N=4.3$ and $7$ using FPGA. The data points were fitted using the allometric equation. To make the fit, we first found the median time for each problem size in Fig.~\ref{fig:2}(a) and (b). A log-log plot for the median times was fitted by a linear curve that was converted into $\sim N^{\alpha}$ dependence for the original variables. We have obtained $\alpha^\textnormal{FPGA}_{4.3}\approx 2.8 \pm 0.5$ and  $\alpha^\textnormal{FPGA}_{7}\approx 1.6 \pm 0.3$. Although the number of data points in our study was limited, the fitting curves show a very reasonable description of the overall tendencies in  Fig.~\ref{fig:2}. 

The FPGA calculations were compared to calculations with the python code on an Intel Core i7-10750H CPU @ 2.60 GHz on the same problem instances. We have observed that the solutions obtained by different methods may differ, and even to reach the same solution the number of integration steps may not be the same. Although in both cases we utilized the same equations, parameters, and initial conditions, there are certain differences in the implementation, such as the precision of floating point numbers. 

\begin{figure}[b]
\centering
    \includegraphics[width=0.48\textwidth]{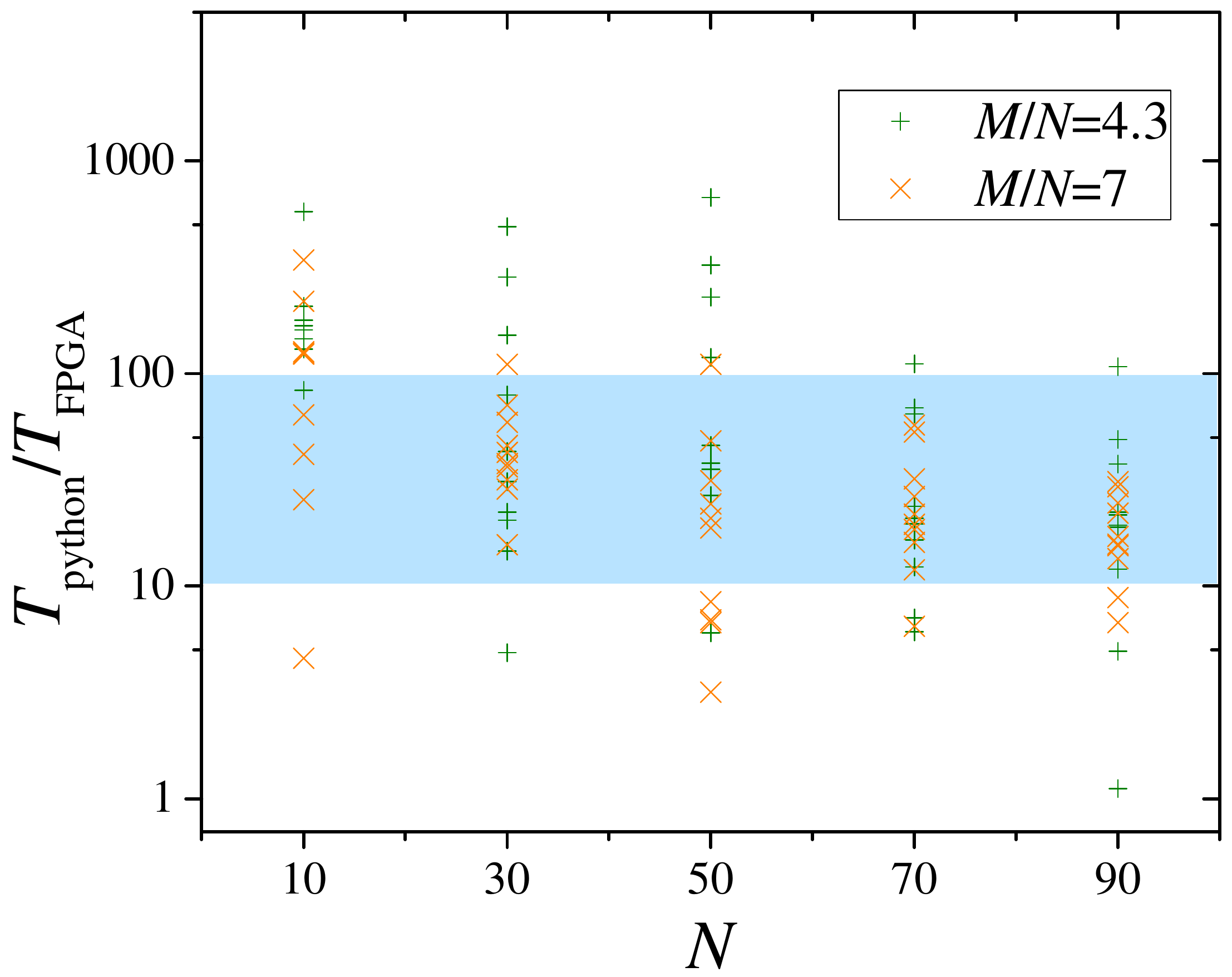}
    \caption{The ratio of the python code calculation time to FPGA calculation time, $T_\textnormal{python}/T_\textnormal{FPGA}$. Each point was obtained for the same problem instance and initial conditions.}
    \label{fig:3}
\end{figure}

To better understand the relationship between the times to reach the solution in the two approaches, in Fig.~\ref{fig:3} we plot the ratio of the python to FPGA times that demonstrates that in all cases the result was found significantly (1-2 orders of magnitude) faster by the FPGA. This plot shows that on average, the ratio $T_\textnormal{python}/T_\textnormal{FPGA}$ slightly decreases with $N$ and it would be interesting to see how it varies with larger $N$. We noticed that in the case of python, the duration of the integration step (in real-time) depends weakly on $N$ and $M/N$. However, no attempts were made to understand this dependence better as well as the efficiency of the python implementation as these topics were not the focus of this study.

\section{Discussion} \label{sec:5}

\begin{figure}[t]
\centering
    \includegraphics[width=0.45\textwidth]{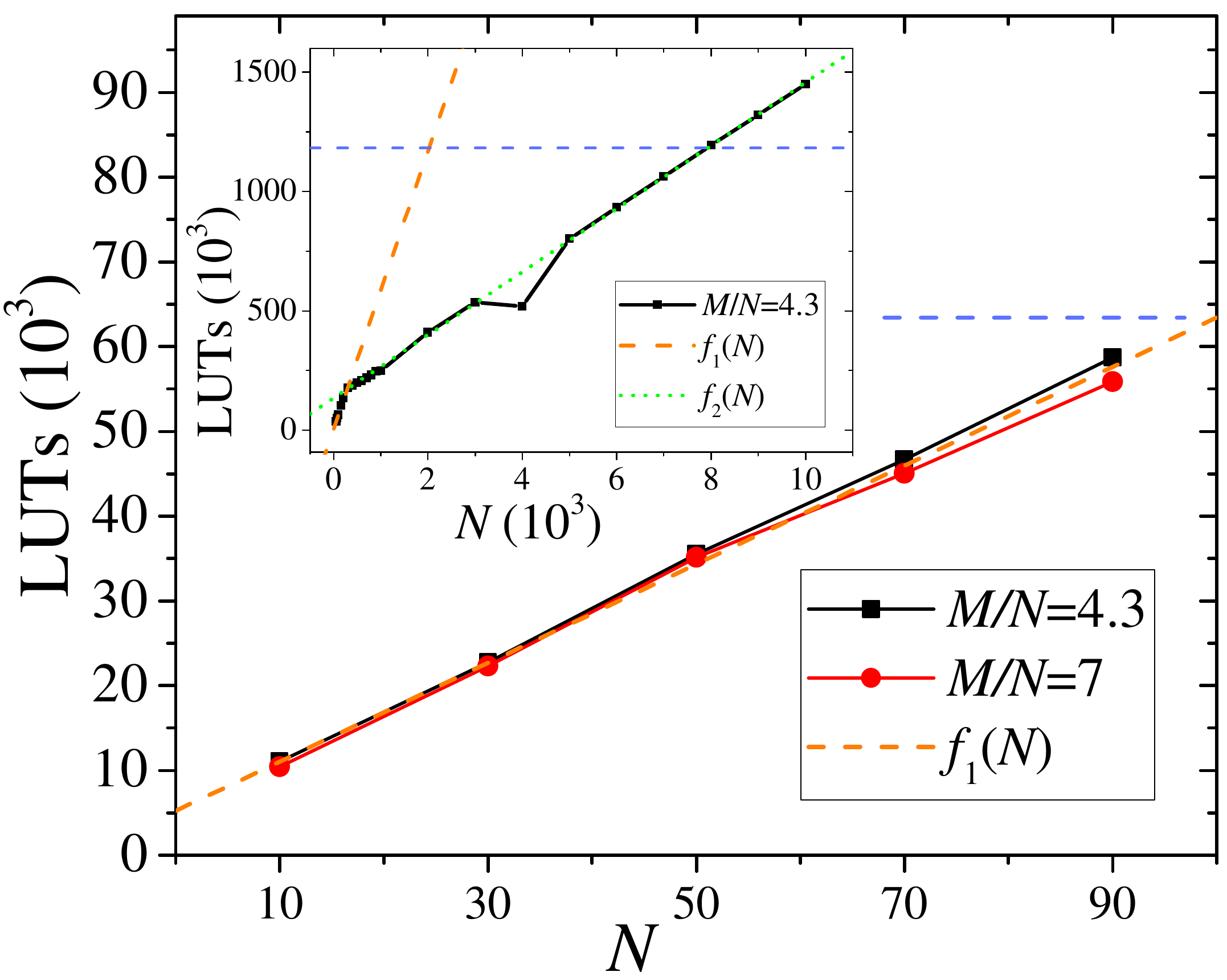}
    \caption{Utilization of LUTs depending on the problem size. The dashed horizontal lines denote the maximum number of LUTs in our small-size FPGA board and VCU118 evaluation board. Inset: LUTs requirement for a larger board obtained using  VIVADO synthesis. The fitting curves are $f_1(N)=5226+582\cdot N$ and $f_2(N)=134147+132\cdot N$.}
    \label{fig:4}
\end{figure}

Our initial results presented here can be considered from several perspectives: the scaling of time to solution, utilization of hardware resources, and comparative performance. 

First of all, we note that the scaling of time to solution (Fig.~\ref{fig:3}) has been estimated using a small set of data points that may not be sufficient to estimate the power-law exponents very accurately. Nonetheless, the power-law exponents we have found with the FPGA are not so different from the ones obtained using a software implementation~\cite{Sean3SAT}. 

To estimate the scaling of the hardware resources, in Fig.~\ref{fig:4} we plot the number of lookup tables (LUTs) in our design versus the number of variables. This plot clearly shows the linear dependence with  582~LUTs per single variable, almost independent of $M/N$. To project this observation to larger boards, we estimated the number of LUTs for a VCU118 evaluation board employing VIVADO synthesis (see the inset in Fig.~\ref{fig:4}).  For smaller values of $N$, we have obtained exactly the same linear dependence as in the case of the small-size board. Interestingly, this initial linear dependence, $f_1(N)$, changes to a slower linear dependence, $f_2(N)$, with an unexpected dip at $N=4000$. Based on this dependence, we expect to be able to fit problems with up to $7.5$k variables into VCU118 based on VU9P device (1,182k LUTs). Note that some currently largest FPGAs, Xilinx VU19P and Intel Stratix 10 GX 10M, contain 4,086k and 10,000k LUTs, respectively, thus potentially increasing the size of the maximum problem that can be solved by an additional order of magnitude.

Finally, it is worth noting from Fig.~\ref{fig:3} that we have obtained a hardware acceleration of the memcomputing solver by one to two orders of magnitude compared to the software. This is consistent with previous hardware accelerations of other types of ODEs~\cite{peccerillo2022survey} and it will be interesting to see if this speed-up holds, or even improves, with larger boards.

\section{Conclusion}\label{concl}

In this work, we have implemented the digital memcomputing algorithm for 3-SAT~\cite{Sean3SAT} on a low-cost FPGA for the first time and performed an initial evaluation of the scalability and resource requirements of such implementation.  Compared to the python simulations, we have observed a significant (1-2 orders of magnitude) reduction in calculation time, although further research is necessary to better understand this trend in larger problem sizes. Our specific method can implement 3-SAT problems with up to tens (and possibly hundreds) of thousands of variables on state-of-the-art FPGA devices. To improve the statistics, a change to the  problem-specific design is desirable\footnote{The present design is instance-specific.}. There are several avenues for further optimization of the present implementation, including the processing of several clauses at a single sub-step that would be interesting to implement on a larger board. 

\newpage

\bibliographystyle{IEEEtran}
\bibliography{SUSYref}

\end{document}